%% file: main.tex
\documentclass[10pt,conference]{IEEEtran}

\input{preamble}

\begin{document}

\title{Tuning LLM-based Code Optimization via Meta-Prompting: An Industrial Perspective}

\author{
\IEEEauthorblockN{
    Jingzhi Gong\IEEEauthorrefmark{5}\IEEEauthorrefmark{2}, 
    Rafail Giavrimis\IEEEauthorrefmark{3}\IEEEauthorrefmark{2}, 
    Paul Brookes\IEEEauthorrefmark{2}, 
    Vardan Voskanyan\IEEEauthorrefmark{2}, 
    Fan Wu\IEEEauthorrefmark{2}, 
    Mari Ashiga\IEEEauthorrefmark{4}\IEEEauthorrefmark{2}, \\
    Matthew Truscott\IEEEauthorrefmark{2}, 
    Mike Basios\IEEEauthorrefmark{2}, 
    Leslie Kanthan\IEEEauthorrefmark{2}, 
    Jie Xu\IEEEauthorrefmark{5}, and
    Zheng Wang\IEEEauthorrefmark{5}$^*$
}
\IEEEauthorblockA{\IEEEauthorrefmark{5}University of Leeds, Leeds, UK}
\IEEEauthorblockA{\IEEEauthorrefmark{2}TurinTech AI, London, UK}
\IEEEauthorblockA{\IEEEauthorrefmark{3}University of Surrey, Surrey, UK}
\IEEEauthorblockA{\IEEEauthorrefmark{4}University of West London, London, UK}
\IEEEauthorblockA{
Emails: \{j.gong, j.xu, z.wang5\}@leeds.ac.uk, mari.ashiga@uwl.ac.uk, \\
\{rafail, paul, vardan, fan, matthew.truscott, mike, leslie\}@turintech.ai
}
\thanks{$^*$Corresponding author: Zheng Wang (z.wang5@leeds.ac.uk).}
}



\maketitle

\input{Sections/abstract}


\input{Sections/introduction}

\input{Sections/background}

\input{Sections/methodology}

\input{Sections/experiment_setup}

\input{Sections/evaluation}

\input{Sections/discussion}



\input{Sections/conclusion}

\bibliographystyle{IEEEtran}
\bibliography{references}

\input{Sections/appendix}

\end{document}

%% file: preamble.tex
\IEEEoverridecommandlockouts

\usepackage{cite}
\usepackage{amsmath,amssymb,amsfonts}
\usepackage{algorithmic}
\usepackage{graphicx}
\usepackage{textcomp}
\usepackage{xcolor}
\usepackage{colortbl}
\usepackage{hyperref}
\usepackage{csquotes}
\usepackage{threeparttable}
\usepackage{rotating}
\usepackage{array}
\usepackage{pifont}
\usepackage{booktabs}
\usepackage{listings}
\usepackage{adjustbox}
\usepackage{multirow}
\usepackage{tcolorbox}
\usepackage{subcaption}
\usepackage{url}
\usepackage{tikz}
\usetikzlibrary{pie}
\usepackage{xspace}
\usepackage[ruled, linesnumbered, vlined, commentsnumbered]{algorithm2e}

\definecolor{beaublue}{rgb}{0.74, 0.83, 0.9}
\definecolor{applegreen}{rgb}{0.55, 0.71, 0.0}
\definecolor{dollarbill}{rgb}{0.52, 0.73, 0.4}
\definecolor{ao(english)}{rgb}{0.0, 0.5, 0.0}
\definecolor{amaranth}{rgb}{0.9, 0.17, 0.31}
\definecolor{cerublue}{rgb}{0.16, 0.32, 0.75}
\definecolor{frenchblue}{rgb}{0.0, 0.45, 0.73}
\definecolor{iceberg}{rgb}{0.44, 0.65, 0.82}
\definecolor{blue-violet}{rgb}{0.24, 0.17, 0.99}
\definecolor{darkorchid}{rgb}{0.6, 0.2, 0.8}
\definecolor{my-violet}{rgb}{0.80,0.79,1.00}
\definecolor{ao}{rgb}{0.0, 0.0, 1.0}

\newcommand{\cmark}{\textcolor{applegreen}{\ding{51}}} 
\newcommand{\xmark}{\textcolor{amaranth}{\ding{55}}} 
\newcommand{\pmark}{\textcolor{frenchblue}{\textbf{$\sim$}}} 

\def\BibTeX{{\rm B\kern-.05em{\sc i\kern-.025em b}\kern-.08em
    T\kern-.1667em\lower.7ex\hbox{E}\kern-.125emX}}

\newcommand{\model}{\textsc{Mpco}\xspace}

\newcommand{\repo}{https://github.com/gjz78910/MPCO}

\newtcolorbox{quotebox}{colback=applegreen!30,boxrule=0.4pt,colframe=black,fonttitle=\bfseries,top=2pt,bottom=2pt}

\newtcolorbox{suggestionbox}{colback=cerublue!20,boxrule=0.4pt,colframe=cerublue,fonttitle=\bfseries,top=2pt,bottom=2pt}

\newcommand{\keybox}[1]{
\begin{tcolorbox}[leftrule=1mm,rightrule=1mm,toprule=0mm,bottomrule=0mm,left=5pt,right=6pt,top=2pt,bottom=2pt, colback=cerublue!30]
\em #1
\end{tcolorbox}
}

\DeclareMathAlphabet\mathbfcal{OMS}{cmsy}{b}{n}

\newcommand{\vect}[1]{\overline{\boldsymbol{#1}}}

%% file: Sections/abstract.tex
\begin{abstract}
    There is a growing interest in leveraging multiple large language models (LLMs) for automated code optimization. However, industrial platforms deploying multiple LLMs face a critical challenge: prompts optimized for one LLM often fail with others, requiring expensive model-specific prompt engineering. This cross-model prompt engineering bottleneck severely limits the practical deployment of multi-LLM systems in production environments. We introduce Meta-Prompted Code Optimization (\model), a framework that automatically generates high-quality, task-specific prompts across diverse LLMs while maintaining industrial efficiency requirements. \model~leverages meta-prompting to dynamically synthesize context-aware optimization prompts by integrating project metadata, task requirements, and LLM-specific contexts.  It is an essential part of the \textsc{Artemis} code optimization platform for automated validation and scaling.
    
    Our comprehensive evaluation on five real-world codebases with 366 hours of runtime benchmarking demonstrates \model's effectiveness: it achieves overall performance improvements up to 19.06\% with the best statistical rank across all systems compared to baseline methods. Analysis shows that 96\% of the top-performing optimizations stem from meaningful edits. Through systematic ablation studies and meta-prompter sensitivity analysis, we identify that comprehensive context integration is essential for effective meta-prompting and that major LLMs can serve effectively as meta-prompters, providing actionable insights for industrial practitioners.
    \end{abstract}

%% file: Sections/introduction.tex
\section{Introduction}
\label{sec:introduction}
Large language models (LLMs) have demonstrated remarkable capabilities in many aspects across the software development lifecycle, including code completion \cite{husein2025large}, natural‑language documentation \cite{dvivedi2024comparative}, test-case generation \cite{jorgensen2024large}, automated debugging \cite{kang2025explainable}, and more \cite{hou2024large}. Beyond these capabilities, recent research studies have demonstrated that LLMs can systematically analyze code patterns to uncover optimization opportunities and produce revised implementations, ranging from memory-management enhancements and algorithmic refinements to automated parallelization and vectorization—thereby enabling high‑performance, resource‑efficient software \cite{gong2025language, ishida2024langpropcodeoptimizationframework,cummins2024meta}.

\begin{figure}[t]
    \centering
    \includegraphics[width=\columnwidth]{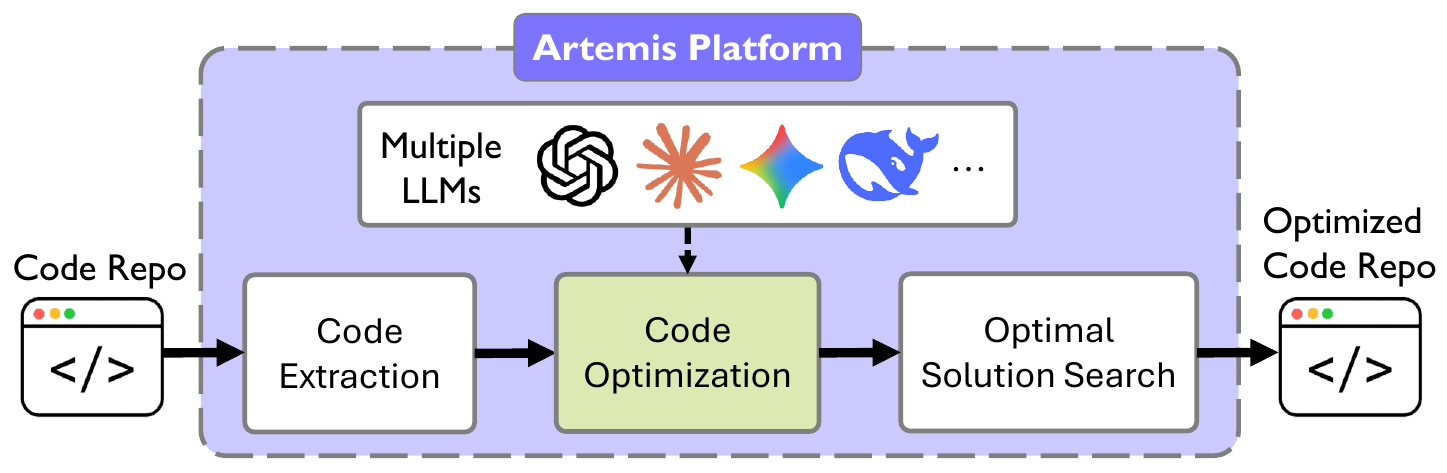}
    \caption{\textsc{Artemis}' multi-LLM code optimization workflow.}
    \label{fig:artemis_workflow}
\end{figure}




  

The success of LLM-based code optimization has led to the development of industrial solutions that integrate these capabilities into production workflows. These platforms provide enterprise-grade infrastructure for deploying LLM-driven code optimization at scale, serving diverse client needs from high-frequency trading systems requiring microsecond latencies to scientific computing applications demanding maximum throughput \cite{aws2020codeguruprofiler2, lin2025eco, artemis_paper, verma2025perfinsights, jiang2025aide}. Among others, TurinTech AI's \textsc{Artemis} platform \cite{artemis_website} represents a leading AI-driven software performance engineering system, which deploys multiple LLMs at scale to search for optimal optimizations of critical code snippets with real-time validation, as illustrated in Figure~\ref{fig:artemis_workflow}. It now processes tens of thousands of requests daily and offers automated performance-engineering solutions that reduce development costs and accelerate time-to-market for performance-critical applications \cite{artemis_businesswire}.

However, in its production deployment, \textsc{Artemis} exposes a fundamental obstacle in multi-LLM code optimization, which we term the ``\emph{cross-model prompt engineering challenge}'' - a prompt that produces excellent optimizations with one LLM may yield poor results with others \cite{sabbatella2024prompt, sahoo2024systematic}. Consequently, platforms need to develop and maintain model-specific prompts for each LLM-task combination; furthermore, industrial platforms often face strict efficiency constraints that limit the feasibility of advanced prompt optimization methods that require extensive trial-and-error or computation \cite{sahoo2024systematic}. This raises the critical question for this work:

\begin{displayquote}
\textit{How can we automatically generate high‑quality, task‑specific prompts across diverse code optimization LLMs while maintaining industrial efficiency requirements?}
\end{displayquote}



Unfortunately, existing solutions cannot adequately address these requirements. Conventional prompting studies like Chain-of-Thought (CoT) focus primarily on general-purpose tasks and single-model scenarios, lacking the specialized contextual awareness and cross-model adaptability required for industrial code optimization \cite{sahoo2024systematic}. Evolutionary prompt optimization methods require pre-defined evaluation functions for each project and multi-round processing, causing scalability problems \cite{chen2023evoprompt, pryzant2023automatic}. Advanced prompt tuning agents, while more flexible and powerful, demand extensive computing resources, impractical for production environments requiring rapid processing \cite{liu2024large, duan2024exploration}.

To address this, we introduce Meta-Prompted Code Optimization (\model), an end-to-end framework that automates prompt engineering while maintaining industrial requirements for flexibility and efficiency across multiple LLMs. In summary, our key contributions are:
\begin{itemize}
    \item We present a novel prompt engineering framework that feeds a meta-prompting LLM rich contextual information to efficiently generate model-adaptive prompts, thereby addressing the cross-model prompting challenge that limits industrial platforms.
    \item We demonstrate \model's effectiveness through comprehensive evaluation across multiple LLMs on the \textsc{Artemis} platform using five real-world codebases and three major LLMs. Each optimization is verified through 10 repeated benchmarking, with total validation time exceeding 336 hours.
    \item Through detailed ablation studies and meta-prompter sensitivity analysis, we identify that comprehensive context integration is essential for effective meta-prompting, and that all three major LLMs can serve effectively as meta-prompters, providing actionable recommendations for industrial deployment.
\end{itemize}


%% file: Sections/background.tex
\input{Tables/comparison_table}

\section{Preliminaries}
\label{sec:background}

\subsection{Industrial Multi-LLM Code Optimization}

Code optimization aims to improve software performance (e.g., making program run fast or reducing the binary size) via algorithmic and low-level code transformations. It has traditionally relied on handcrafted algorithmic tweaks, compiler-driven pre-defined transformations (e.g., loop unrolling, inlining, and dead-code elimination), and machine learning–based autotuning systems \cite{wang2018machine}. Yet, these methods often struggle to capture domain-specific performance patterns, generalize across diverse codebases, and incur high search overheads~\cite{wang2018machine, cummins2023large}. Recent advances in LLMs enable automated code optimization systems that learn from vast code corpora and leverage both a static knowledge base and environment interaction to generate adaptive, context‑aware optimization suggestions, without expert intervention or limitation on pre-defined operations \cite{gong2025language}. Today, industrial platforms integrate these LLM-driven optimizations into production pipelines, offering low‑latency inference and high-throughput processing \cite{aws2020codeguruprofiler2, artemis_website}.

However, single-LLM approaches cannot meet these diverse industrial requirements effectively, as different models exhibit complementary strengths that are needed for comprehensive optimization coverage \cite{artemis_paper}. Consequently, industrial platforms like \textsc{Artemis} deploy multiple LLMs simultaneously to leverage their diverse capabilities \cite{artemis_paper,artemis_businesswire}. Specifically, Figure~\ref{fig:artemis_workflow} illustrates the \textsc{Artemis} workflow, where the codebase is first analyzed to identify optimization targets, then sent to a diverse set of LLMs for improvement suggestions, and finally filtered through an intelligent search to find the optimal solution.

\begin{figure}[t]
\centering



  
\includegraphics[width=\columnwidth]{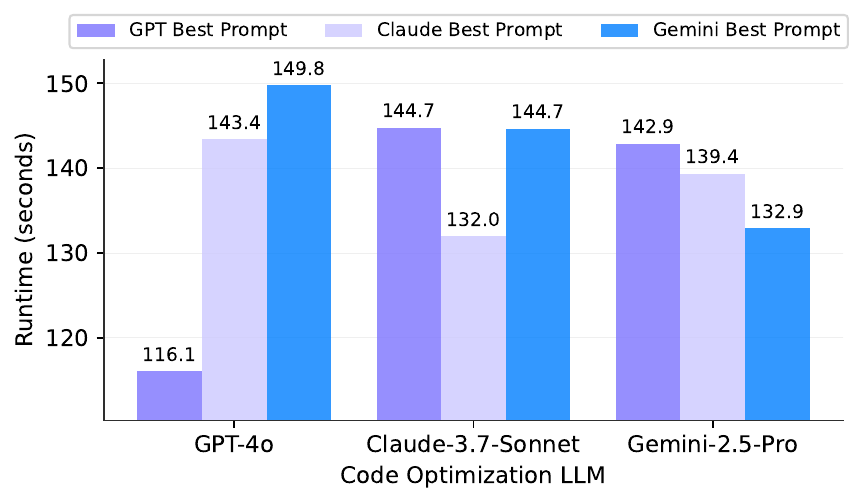}
\caption{Cross-model optimization results on \textsc{Llama.cpp}.}
\label{fig:challenge_example}
\end{figure}

\subsection{Cross-Model Prompt Engineering Bottleneck}
The effectiveness of LLM‑generated optimizations depends critically on prompt quality \cite{sabbatella2024prompt, sahoo2024systematic}, yet different models exhibit varying sensitivities to prompt structure, context, and optimization strategies. As a motivating example, consider optimizing a
hot code region of the \textsc{Llama.cpp} inference framework~\cite{ggerganov_llama_cpp}. Figure~\ref{fig:challenge_example} shows that optimizing the same code snippet with three LLMs using different optimal prompt templates reveals significant performance variations. Especially, GPT-4o achieves its best performance (116.1s) with its own prompt template, but degrades by 23.5\% (143.4s) when using Claude's template and 29.0\% (149.8s) with Gemini's template. These inconsistencies often stem from model-specific factors - differences in training corpora and architectures, variations in tokenization and context-window handling, and divergent decoding defaults and instruction‑following behaviors~\cite{liu2023pre, zhou2022large, sahoo2024systematic}.

To achieve reliable performance across multiple LLMs, platforms often have to optimize and maintain distinct prompts for every LLM-task-project pairing. The resulting multiplicative complexity renders traditional prompting workflows infeasible in production environments where rapid, large‑scale, generalizable optimizations are required.

This challenge is uniquely severe in code optimization compared to generic natural language processing (NLP) tasks. Unlike text generation, where outputs can be stylistically varied, optimized code has strict, non-negotiable correctness requirements: it must compile, pass all functional tests, and be free of runtime side effects like memory leaks. The determinism and reproducibility of optimizations are paramount. A slight, seemingly innocuous change in a prompt can lead to a syntactically correct but functionally flawed or slower implementation, making the cross-model challenge a critical barrier to reliable industrial deployment.

\subsection{Existing Prompt Optimization Approaches and Limitations}
Several approaches have been proposed to automate prompt engineering, but each exhibits critical limitations when applied to multi-LLM industrial code optimization scenarios, as summarized in Table~\ref{tab:approach_comparison}.

\textbf{Baseline prompting techniques} include the most common prompt designs such as direct instructions (simple and ad-hoc task specifications), Chain-of-Thought reasoning (step-by-step problem decomposition) \cite{wei2022chain}, few-shot learning (task demonstration through examples) \cite{DBLP:conf/iclr/ShypulaMZ0GYHNR24}, and contextual prompting (fixed template with rich task-related contexts) \cite{huang2024effilearner}. While efficient and easy to deploy, these methods rely on static, manually-crafted templates that cannot adapt seamlessly across different optimization metrics, models, or project types.

\textbf{Evolutionary prompting methods} encompass both automatic prompt optimization approaches such as EvoPrompt (genetic algorithm-based prompt evolution) \cite{chen2023evoprompt} and APE (automatic prompt engineering through iterative refinement) \cite{zhou2022large}, as well as gradient-based prompt tuning techniques (continuous optimization of prompt embeddings) \cite{pryzant2023automatic, chen2024reprompt, sabbatella2024prompt}. While these methods achieve substantial gains for well-defined tasks, they require multiple rounds of LLM interaction, leading to high computation overhead \cite{chen2023evoprompt}, which could be infeasible for platforms processing tens of thousands of daily requests \cite{artemis_businesswire}. Furthermore, they often rely on manual design of evaluation pipelines for each task/project, limiting their scalability in production environments.

\textbf{Agentic prompting systems} employ autonomous AI agents that iteratively refine prompts through environmental interaction, feedback incorporation, and self-reflection mechanisms \cite{wang2023promptagent, zhang2025mars}. These agent-based approaches demonstrate context-awareness and cross-domain adaptability by dynamically adjusting prompts based on task requirements and model responses. However, they usually incur higher deployment complexity and overhead due to extensive environment interaction cycles, action planning, agent state updates, and memory I/O \cite{wang2025ai}, making them unsuitable for large-scale industrial scenarios requiring rapid responses.

\textbf{Existing meta-prompting approaches} \cite{zhang2023meta,madaan2023self,gong2024self,hou2022metaprompting, suzgun2024meta} employ higher-order prompts to autonomously generate task-specific instructions, and have shown promise in domains such as text classification \cite{hou2022metaprompting}, dialog response generation \cite{suzgun2024meta, madaan2023self}, and mathematical reasoning \cite{zhang2023meta}. However, these approaches may either generate generic prompts without context-awareness, lack end-to-end deployment efficiency, or have not been validated on real-world industrial environments.

%% file: Tables/comparison_table.tex
\begin{table*}[t]
    \centering
    \caption{Comparison of prompting approaches. \cmark = Full support, \pmark = Partial support, \xmark = No support}
    \label{tab:approach_comparison}
    \begin{tabular}{>{\centering\arraybackslash}p{2.8cm}>{\centering\arraybackslash}p{1.5cm}>{\centering\arraybackslash}p{1.6cm}>{\centering\arraybackslash}p{1.6cm}>{\centering\arraybackslash}p{1.5cm}>{\centering\arraybackslash}p{1.6cm}>{\centering\arraybackslash}p{1.6cm}>{\centering\arraybackslash}p{1.7cm}}
    \toprule
\textbf{Approach} & \textbf{Context-Awareness} & \textbf{End-to-End Automation} & \textbf{Deployment Simplicity} & \textbf{Low Overhead} & \textbf{Cross-Metric Adaptability} & \textbf{Cross-Model Adaptability} & \textbf{Cross-Project Adaptability} \\ \hline
Manual Prompting & \pmark & \xmark & \cmark & \xmark & \xmark & \xmark & \xmark \\
Baseline Prompting & \xmark & \cmark & \cmark & \cmark & \xmark & \xmark & \xmark \\
Evolutionary Prompting & \xmark & \cmark & \pmark & \xmark & \pmark & \pmark & \pmark \\
Agentic Systems & \cmark & \cmark & \xmark & \xmark & \cmark & \cmark & \cmark \\
\rowcolor{applegreen!20} \model~(Our Approach) & \cmark & \cmark & \cmark & \cmark & \cmark & \cmark & \cmark \\ 
    \bottomrule
    \end{tabular}
\end{table*} 

%% file: Sections/methodology.tex
\section{The \model~Framework}
\label{sec:methodology}


To address these limitations, we propose Meta-Prompted Code Optimization (\model), a meta-prompting framework specifically designed for multi-LLM industrial code optimization. As summarized in Table \ref{tab:approach_comparison}, \model~differs from existing approaches by systematically incorporating task-specific contexts, providing efficient and scalable deployment, and integrating with automated performance 
validation requiring minimal human intervention.

\begin{figure}[t!]
    \centering
    \includegraphics[width=\columnwidth]{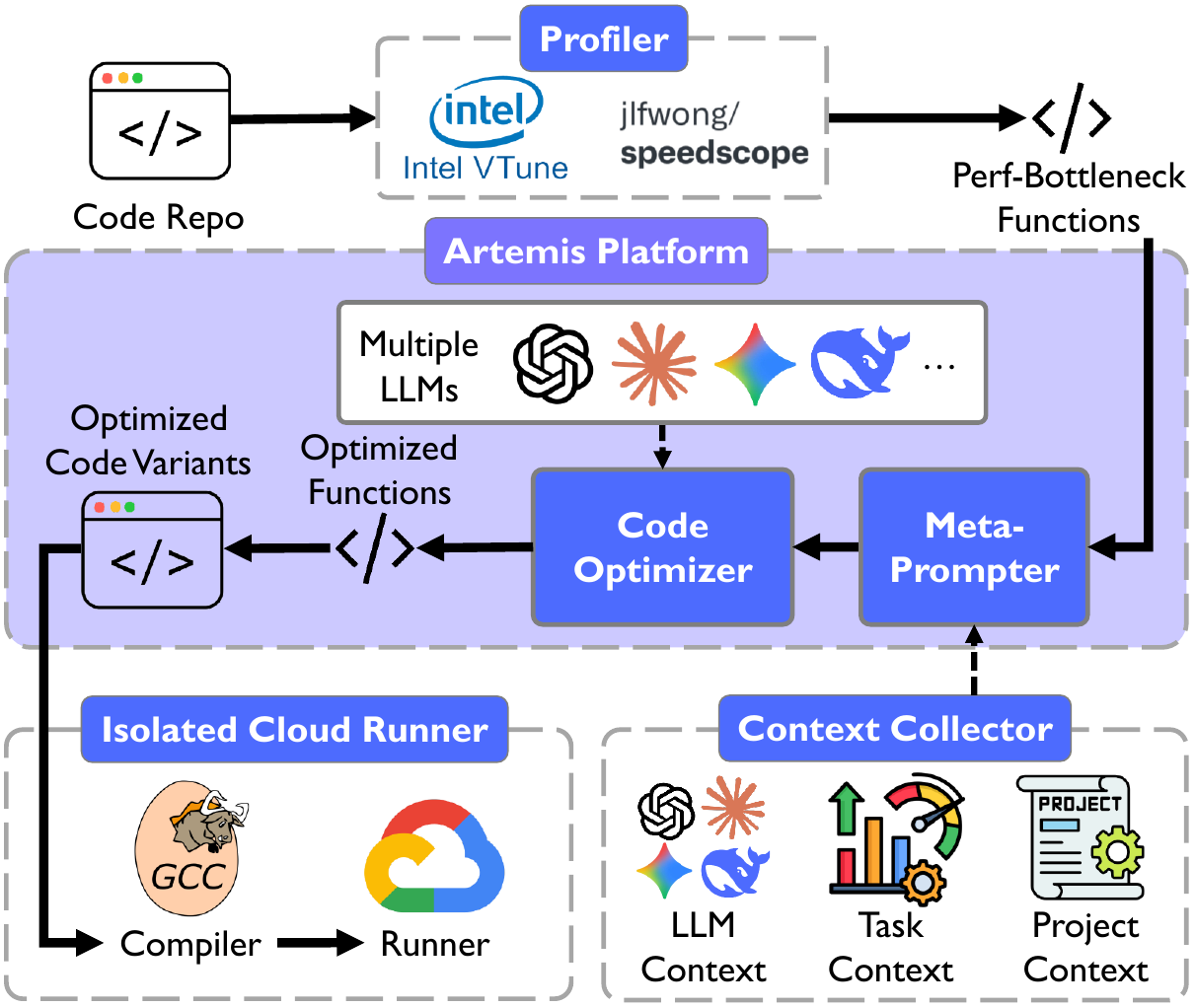} 
    \caption{The end-to-end workflow of \model.}
    \label{fig:workflow}
\end{figure}

The architecture of \model~(Figure \ref{fig:workflow}) consists of four core stages that ensure complete automation from profiling to validated optimization results, as delineated below.

\subsection{Stage 1: Profiling and Bottleneck Identification}
The framework begins by systematically identifying performance bottlenecks within real-world software projects. In this stage, users specify software projects requiring optimization, and the framework applies industry-standard profilers - Intel VTune Profiler \cite{intel2021vtune} for C++ and Speedscope \cite{speedscope2020} for Python - to locate performance-critical code snippets. These profiling tools use instrumentation to trace function calls and resource usage at runtime, and visualize hotspots via flame graphs or detailed reports. We opt for VTune and Speedscope in this study because VTune provides low‑overhead, cycle‑accurate profiling of C++ code \cite{patsnap2025cpuprofiling}, while Speedscope offers rapid and efficient exploration for dynamic Python workloads \cite{hanselman2020speedscope}.

We then rank the identified performance bottlenecks by their impact on runtime and forward the top-10 code snippets to the next stage, enabling our framework to prioritize efficiency-critical code rather than arbitrary selections. Formally, this profiling process can be defined as:

\begin{equation}
\vect{\mathbfcal{B}} = \texttt{TopK}(\texttt{Profile}(\mathbfcal{R}), k=10),
\end{equation}

where $\mathbfcal{R}$ is the input code repository, $\texttt{Profile}$ is the profile function, $\texttt{TopK}$ selects the $k$ functions with the highest performance impact, and $\vect{\mathbfcal{B}} = \{\mathbfcal{B}_1, \mathbfcal{B}_2, \ldots, \mathbfcal{B}_{10}\}$ is the set of top-10 bottleneck code snippets selected for optimization.

\subsection{Stage 2: Context Collection and Meta-Prompt Generation}

To overcome the cross‑model prompt engineering bottleneck, \model~introduces a novel meta‑prompting mechanism that swiftly collects and unifies comprehensive relevant context for prompt synthesis. Unlike prior works that either use generic templates without context awareness \cite{suzgun2024meta, zhang2023meta} or incur large overhead in context collection \cite{chen2023evoprompt, zhang2025mars}, \model~builds on Artemis' existing metadata store to enable rapid context gathering through three lightweight contexts:

\begin{enumerate}
  \item \textbf{Project context} extracts project metadata (name, description, and languages) from Artemis's existing project database, ensuring the generated prompt maintains syntactic validity and project consistency.
  \item \textbf{Task context} maintains optimization-specific requirements (target metrics, domain constraints, and optimization considerations) in a simple JSON format that updates dynamically as new performance goals emerge.
  \item \textbf{LLM context} stores model-specific characteristics (target LLM, model strengths and limitations) in database tables that scale with new model deployments.
\end{enumerate}

This Artemis-integrated context storage enables context collection in seconds, without the computational overhead of traditional and more advanced prompt optimization methods, and also enables easy maintenance and updates, allowing the framework to scale up seamlessly as the platform evolves with new model deployments and optimization requirements.

The meta-prompting LLM receives structured contexts through the \model~template (Figure~\ref{fig:mpco_template}) and produces specialized prompts in a single inference call, eliminating the need for per-model manual tuning while maintaining industrial-grade efficiency. Formally, this process can be defined as a two-step process:

\begin{equation}
\mathbfcal{P}_{m,t,p} = \texttt{GenPrompt}(\mathbfcal{T}(\mathbfcal{C}_{m}, \mathbfcal{C}_{t}, \mathbfcal{C}_{p})),
\end{equation}

where $\mathbfcal{C}_{m}, \mathbfcal{C}_{t}, \mathbfcal{C}_{p}$ represent model, task, and project contexts respectively, $\mathbfcal{T}$ is the meta-prompt template, \texttt{GenPrompt} is the meta-prompting function, and $\mathbfcal{P}_{m,t,p}$ is the generated prompt for model $m$, task $t$, and project $p$.

\input{Figures/meta_prompt_template}

\subsection{Stage 3: Multi-LLM Code Optimization}  
In this stage, generated meta-prompts are deployed across multiple target LLMs simultaneously to leverage 
diverse optimization capabilities \cite{artemis_paper}. For each bottleneck function $\mathbfcal{B}_i$ identified in Stage 1, each LLM receives its model-specific prompt along with the performance-critical code snippet, producing code optimizations that reflect both the model's inherent strengths and the contextual guidance provided by the meta-prompt.

For each optimized code snippet, the framework then generates a variant of the original code repository where only one specific code snippet is optimized, while preserving the complete project context including build configurations, dependency relationships, and test suites. This ensures that the optimized code can be validated under identical conditions to the original implementation, addressing a critical requirement for performance evaluation.

The optimization process is implemented within the Artemis platform, which orchestrates the multi-LLM deployment and manages the code variant generation workflow. Formally, the process can be defined as:

\begin{equation}
\mathbfcal{O}_i = \texttt{LLM}(\mathbfcal{P}_{m,t,p}, \mathbfcal{B}_i),
\end{equation}

\begin{equation}
\mathbfcal{V}_i = \texttt{GenVariant}(\mathbfcal{R}, \mathbfcal{O}_i),
\end{equation}

where $\mathbfcal{B}_i$ is a bottleneck code snippet, $\mathbfcal{P}_{m,t,p}$ is the generated prompt, \texttt{LLM} is the code optimization function, $\mathbfcal{R}$ is the original code repository, $\texttt{GenVariant}$ is the code variant generation function, $\mathbfcal{O}_i$ and $\mathbfcal{V}_i$ are the optimized code snippet and code repository variant for bottleneck $\mathbfcal{B}_i$, respectively.

\subsection{Stage 4: Performance Evaluation and Validation}
Finally, the framework automatically validates the generated code variants through isolated cloud services provided by \textsc{Artemis}. The validation process includes: compiling all code variants to ensure syntactic correctness, executing comprehensive unit test suites to verify functional equivalence, running performance benchmarks, collecting detailed runtime metrics, and statistical significance analysis.

This systematic validation ensures that only functionally correct and performance-validated optimizations are accepted, maintaining the reliability standards required for industrial deployment. Formally, this process can be defined as:

\begin{equation}
\mathbfcal{E}_{i} = \texttt{Validate}(\mathbfcal{V}_{i}),
\end{equation}

where $\mathbfcal{V}_{i}$ is a code variant from Stage 3, $\texttt{Validate}$ is the function that performs compilation, unit testing, and benchmarking, and $\mathbfcal{E}_{i}$ is the evaluated performance result.

%% file: Figures/meta_prompt_template.tex
\begin{figure}[t]
    \centering
    \begin{lstlisting}[basicstyle=\footnotesize\ttfamily, frame=single, breaklines=true, xleftmargin=0.2cm, xrightmargin=0.2cm, moredelim={[is][\color{blue}]{@}{@}}, moredelim={[is][\color{applegreen}]{§}{§}}, moredelim={[is][\color{frenchblue}]{¤}{¤}}, moredelim={[is][\color{amaranth}]{€}{€}}]
You are an expert in code optimization. Please generate a prompt that will instruct the target LLM @{target_llm}@ to optimize code for @{objective}@. 
Consider the project context, task context, and adapt the prompt complexity and style based on the target LLM's capabilities.

§## Project Context
Project Name: {project_name}
Project Description: {project_description}
Primary Languages: {project_languages}§

¤## Task Context
- Description: {task_description}
- Considerations: {task_considerations}¤

€## Target LLM Context
- Target Model: {target_llm}
- Considerations: {llm_considerations}€

    \end{lstlisting}
    \caption{The \model~meta-prompting template.}
    \label{fig:mpco_template}
\end{figure}

%% file: Sections/experiment_setup.tex
\section{Experiment Setup}
\label{sec:experiment_setup}

\subsection{Research Questions}
To evaluate the effectiveness of \model, we conducted comprehensive experiments to answer the following research questions (RQs):
\begin{itemize}
    \item \textbf{RQ1}: How does \model~address the cross-model prompt engineering bottleneck compared to baseline prompting approaches?
    \item \textbf{RQ2}: What aspects of \model's context collection most significantly impact its effectiveness?
    \item \textbf{RQ3}: How sensitive is \model~to the choice of meta-prompting LLM?
\end{itemize}

\subsection{Subject Systems and Bottleneck Selection}
We systematically selected five subject systems from open-source GitHub repositories: we identified candidate repositories through GitHub searches for PRs labeled ``\emph{runtime optimization}'', ``\emph{speedup}'', or ``\emph{runtime improvement}'' \cite{kalliamvakou2014promises}, and filtered them by requiring built-in benchmarks, clear build systems, and target language support. Our study concentrated on C++ and Python projects, as these languages are dominant in open-source performance-critical systems and are widely supported by state-of-the-art LLMs and profiling tools \cite{tiobe2025, intel2021vtune, speedscope2020}. For each project, we selected the top-10 most performance-critical code snippets based on the profiling data. 

This selection strategy ensures that our evaluation covers the most relevant industrial scenarios where performance optimization is critical \cite{kalliamvakou2014promises}. Table~\ref{tab:subject_systems} summarizes the selected systems, including project name, description, language, and evaluation benchmark.

\input{Tables/subject_systems_table}

\subsection{Performance Evaluation and Validation}
Each generated code optimization was integrated into a new code variant, and each variant was submitted to Artemis for automated compilation and execution. Any LLM response that failed to adhere to the required format (containing only the synthesized prompt/optimized code without additional commentary) was excluded from analysis. Similarly, code optimizations that resulted in compilation errors or runtime failures were systematically removed from the dataset prior to statistical analysis. This filtering process ensures that our performance comparisons are based solely on functionally correct and executable code variants, providing a fair and meaningful evaluation of optimization effectiveness across different prompting approaches. 

To control measurement noise, we isolated the experiment environment and executed each benchmark 10 times, ensuring statistical significance and stable performance distributions. The LLMs used were GPT-4o (OpenAI) \cite{openai2024gpt4o}, Gemini 2.5 Pro (Google) \cite{kavukcuoglu2025gemini2.5}, and Claude 3.7 Sonnet (Anthropic) \cite{anthropic2025claude3.7}, selected to represent the three major LLM providers and diverse architectural approaches. We used their default temperature settings to ensure reproducible results and avoid introducing additional variance from configuration tuning.

All experiments were conducted on a dedicated Google Cloud Platform server with 4 CPUs, 16GB RAM, running Ubuntu 25.04. The benchmarks were executed in Docker containers to ensure consistent execution environments and eliminate interference between different optimization runs. Each experiment was executed in an isolated environment without any other processes running to minimize measurement noise and ensure reproducible results.

\subsection{Evaluation Metric and Statistical Analysis}
To evaluate each code optimization, we adopt the most common metric in the LLM-based code optimization literature \cite{gong2025language}, i.e., the percentage runtime performance improvement (\%PI), computed as:
\begin{equation}
\%PI \;=\;\frac{T_{{orig}} - T_{{opt}}}{T_{{orig}}}\;\times 100\%\,, 
\end{equation}
where $T_{{orig}}$ is the mean runtime of the original code and $T_{{opt}}$ is the mean runtime of the optimized code version.

We further conduct rigorous statistical analysis to rank prompt variants using Mann-Whitney U test \cite{mann1947test} and Cohen's d effect size \cite{cohen1988statistical}. In particular, the Mann-Whitney U test—an assumption‐free, non-parametric method suited to our independent, continuous \%PI measurements—is applied to detect shifts in central tendency between optimization approaches, and Cohen’s d is computed to standardize the observed mean differences and confirm practical relevance. 

Our ranking strategy works as follows: (1) calculate the mean \%PI for each approach, (2) sort approaches by mean \%PI in descending order, and (3) assign ranks sequentially---for each approach, if it is statistically significantly different from the previous approach ($p \leq 0.05$ or $|d| \geq 0.2$)\footnote{The $0.05$ significance level follows established norms in SE research \cite{arcuri2011practical}, while the $0.2$ threshold for effect size ensures that differences considered are not only statistically significant but also practically relevant \cite{sullivan2012using}.}, assign it the next rank; otherwise, assign it the same rank as the previous approach. By doing so, we group similar-performing methods together while distinguishing methods with meaningful performance differences.

%% file: Tables/subject_systems_table.tex
\begin{table}[t]
    \centering
    \setlength{\tabcolsep}{0.6mm}
    \renewcommand\arraystretch{1.1}
    \caption{Summary of subject systems.}
    \label{tab:subject_systems}
    \begin{adjustbox}{width=\columnwidth,center}
    \begin{tabular}{l l l l}
        \toprule
        \textbf{System} & \textbf{Description} & \textbf{Lang.} & \textbf{Benchmark} \\
        \midrule
        \href{https://github.com/paulsbrookes/BitmapPlusPlus}{BitmapPlusPlus} & Single‑header BMP image library & C++ & Built-in chess\_board bench \\
        \href{https://github.com/ggerganov/llama.cpp}{Llama.cpp}      & Efficient LLM inference engine    & C++ & Built-in bench with TinyLlama \\
        \href{https://github.com/RPCS3/rpcs3}{RPCS3}                 & PlayStation 3 emulator            & C++ & Built-in rpcs3\_benchmark \\
        \href{https://github.com/openai/whisper}{Faster‑Whisper}     & Fast speech‑to‑text transcription & Python & Built-in transcription bench \\
        \href{https://github.com/langflow/langflow}{Langflow}        & Visual flow‑based LLM workflows   & Python & Built-in benchmark\_operations \\
        \bottomrule
    \end{tabular}
    \end{adjustbox}
\end{table}

%% file: Sections/evaluation.tex
\section{Evaluation}
\label{sec:evaluation}
This section presents the results of our experiment, structured around the research questions defined in Section \ref{sec:experiment_setup}.

\subsection{RQ1: \model~Effectiveness vs. Baseline Prompting}
\subsubsection{Method}
In industrial platforms, code optimizations are often expected to complete within seconds and at minimal computational cost. The advanced prompting techniques discussed in Section~\ref{sec:background}, which often rely on extensive reasoning chains or agent coordination \cite{chen2023evoprompt, wang2023promptagent}, are thus beyond the scope of this study. Instead, we benchmarked \model~against established baseline prompting methods most commonly used in LLM-based code optimization \cite{gong2025language}\footnote{The detailed implementation of all baseline prompts is available \href{\repo}{here}.}, including:

\begin{itemize}
    \item \textbf{Chain-of-Thought (CoT):} Structured  prompts that guide the LLM through step-by-step optimization \cite{wei2022chain}.

    \item \textbf{Few-Shot Prompting:} Prompts incorporating optimization examples to demonstrate desired output patterns \cite{DBLP:conf/iclr/ShypulaMZ0GYHNR24}.

    \item \textbf{Contextual Prompting:} Prompts with the same context information as \model~but without meta-prompting \cite{huang2024effilearner}.
\end{itemize}

The evaluation procedures followed the methodology detailed in Section~\ref{sec:experiment_setup}. Each method was applied to optimize all bottleneck code snippets across the five subject systems (50 in total) and across three target LLMs; for each optimization, we generated a code variant, measured its runtime through 10 benchmark executions (resulting in up to 1500 measurements per method), and conducted statistical tests (Mann-Whitney U test and Cohen's d effect size) to rank different approaches.

\input{Tables/RQ1_results}

\subsubsection{Results}
Table~\ref{tab:rq1_results} presents the results for \textbf{RQ1}. Clearly, \model~achieves the best average rank (1.00) for all subject systems, significantly outperforming all baseline methods and demonstrating consistent gains across the three LLMs used for code optimization.

In particular, \model~yielded the highest \%PI in four out of five systems, with strong results on \textsc{BitmapPlusPlus} (19.06\%), \textsc{Langflow} (9.01\%), and \textsc{Llama.cpp} (7.84\%). Notably, on \textsc{BitmapPlusPlus}, both CoT and Few-shot prompting performed similarly to \model, and all three methods shared rank 1. This is likely due to the relatively standard and simple nature of the performance issues in \textsc{BitmapPlusPlus}, which can be effectively addressed by the predefined patterns in few-shot learning prompts or by the reasoning capabilities introduced by CoT.

Interestingly, we can see that even when the same contexts were provided, contextual prompting did not achieve competitive results as \model. This highlights the strength of meta-prompting, which dynamically adjusts its guidance based on the specific task and the capabilities of the LLM. In contrast, contextual prompting may overwhelm the LLM with too much information, leading to degraded performance - similar to a teacher who uses the same teaching style for all students, regardless of their individual characteristics. In summary:

\begin{quotebox}
   \noindent
   \textit{\textbf{RQ1:} \model~consistently outperforms baseline prompting methods, achieving the top rank across all systems and the highest mean \%PI in 4/5 systems, demonstrating that its context-aware meta-prompting strategy can effectively address the cross-model prompt engineering challenge.}  
\end{quotebox}

\subsection{RQ2: Ablation Analysis of Contextual Components}
\label{subsec:rq2}
\subsubsection{Method}
To assess the contribution of each contextual component in \model's template, we conducted a systematic ablation analysis by comparing the full template (Figure~\ref{fig:mpco_template}) against three reduced variants:

\begin{itemize}
    \item \textbf{\model$_{\texttt{NP}}$}: No project context (project name, description, and primary languages).
    \item \textbf{\model$_{\texttt{NT}}$}: No task context (task description and optimization considerations).
    \item \textbf{\model$_{\texttt{NL}}$}: No LLM context (model-specific characteristics and adaptation instructions).
\end{itemize}

The evaluation followed the same methodology as \textbf{RQ1}, applying each ablated version to all 50 code snippets across the five subject systems and three target LLMs.

\input{Tables/RQ2_results}

\subsubsection{Results}
Table~\ref{tab:rq2_results} highlights the importance of using comprehensive contexts to \model's effectiveness, where the full \model~template is consistently ranked the first across all systems and gets the highest \%PI in three out of five, with performance gains up to 9.01\%. Even in the two cases where the full \model~template did not yield the highest \%PI - \textsc{Faster‑Whisper} and \textsc{BitmapPlusPlus} - \model~still achieved rank 1 based on statistical tests. These results demonstrate that removing any contextual component noticeably reduces \model's overall effectiveness, underscoring the importance of comprehensive context integration.

For \textsc{Faster‑Whisper}, \model$_{\texttt{NL}}$ slightly outperformed the full template (5.73\% vs. 5.64\%), possibly because the optimization targets in this project are relatively well-known, making them less sensitive to model-specific instructions. In \textsc{BitmapPlusPlus}, all variants, including the full template and ablated versions, shared the top rank, suggesting that optimization challenges in this system are relatively straightforward, and even limited contextual guidance is sufficient, which aligns with our findings from \textbf{RQ1} that performance issues in \textsc{BitmapPlusPlus} appeared easier to diagnose.

Interestingly, the average ranks of the ablated variants were similar (2.20, 2.00, and 2.00), indicating that no single contextual component dominates in importance, but rather that their joint combination provides the most consistent benefit. Based on these observations, we have:

\begin{quotebox}
   \noindent
   \textit{\textbf{RQ2:} The full \model~template yields the highest \%PI in 3/5 systems and best average rank (1.00), demonstrating that comprehensive context integration is essential for effective meta-prompting.}  
\end{quotebox}

\subsection{RQ3: Sensitivity Analysis of Meta-Prompting LLM}
\label{subsec:rq3}
\subsubsection{Method}
To answer this, we investigated \model's sensitivity to the choice of meta-prompting LLM by evaluating three different LLM configurations:

\begin{itemize}
    \item \textbf{\model$_{\texttt{4o}}$}: Use GPT-4o as the meta-prompter.
    \item \textbf{\model$_{\texttt{37}}$}: Use Claude 3.7 Sonnet as the meta-prompter.
    \item \textbf{\model$_{\texttt{25}}$}: Use Gemini 2.5 Pro as the meta-prompter.
\end{itemize}

Unlike \textbf{RQ1}, where all methods shared the same meta-prompter (GPT-4o), this RQ evaluated three different meta-prompting LLMs, but all code optimizations were performed by the same optimizer (Claude 3.7 Sonnet) to ensure a controlled comparison.

\subsubsection{Results}
Table~\ref{tab:rq3_results} presents the analysis regarding meta-prompting LLM choice. Notably, \model$_{\texttt{4o}}$ achieved the best average rank (1.00) across all systems and the highest \%PI in three of them, with performance improvements up to 19.77\%. This demonstrates that GPT-4o as a meta-prompter can generate highly effective optimization prompts. 

Similarly, \model$_{\texttt{25}}$ maintained a competitive average rank of 1.20 (rank 1 in four out of five systems), with the best mean \%PI in two systems, suggesting that Gemini 2.5 Pro may be better suited for certain optimization scenarios, likely due to its specialized pretraining and reasoning capabilities.

These findings have practical implications for industrial deployment: while GPT-4o shows a slight performance advantage, all three meta-prompting LLMs provide significant improvements over the original code. Therefore, organizations can confidently choose their meta-prompting LLM based on practical considerations, such as cost, availability, and platform compatibility, without significantly compromising \model's effectiveness. Consequently, we conclude:

\begin{quotebox}
   \noindent
   \textit{\textbf{RQ3:} \model$_{\texttt{4o}}$ ranks best across all systems with with the highest \%PI in 3/5 systems. However, while GPT-4o offers a slight advantage, all three models yield significant gains, confirming that \model's benefits are not tied to a specific meta-prompter.}
\end{quotebox}

\input{Tables/RQ3_results}

%% file: Tables/RQ1_results.tex
\begin{table}[t!]
    \footnotesize
    \centering
    \caption{The mean and standard deviation of \%PI, denoted as Mean (SD), for \model~and the basic prompting approach across three LLMs and five projects. For each case, \setlength{\fboxsep}{1.5pt}\colorbox{applegreen!30}{green cells} mean \model~has the best mean \%PI; or \setlength{\fboxsep}{1.5pt}\colorbox{red!20}{red cells} otherwise. The one(s) with the best rank ($r$) from the statistical tests is in bold.}
    \label{tab:rq1_results}
    \setlength{\tabcolsep}{0.6mm}
    \begin{adjustbox}{width=\columnwidth,center}
    \begin{tabular}{lllllllll}
    \toprule
    \multirow{2}{*}{\textbf{System}} & \multicolumn{2}{c}{\textbf{\model}} & \multicolumn{2}{c}{\textbf{CoT}} & \multicolumn{2}{c}{\textbf{Few-shot}} & \multicolumn{2}{c}{\textbf{Contextual}} \\ \cline{2-9}
 & \textbf{$r$} & \textbf{Mean (SD)} & \textbf{$r$} & \textbf{Mean (SD)} & \textbf{$r$} & \textbf{Mean (SD)} & \textbf{$r$} & \textbf{Mean (SD)} \\ \hline
    \textsc{BitmapPlusPlus} & \textbf{1} & \textbf{19.06} (1.48) & \textbf{1} & \textbf{19.01} (1.53) & \cellcolor{red!20}\textbf{1} & \cellcolor{red!20}\textbf{19.29} (2.37) & 2 & 18.49 (1.17) \\
    \textsc{Llama.cpp} & \cellcolor{applegreen!30}\textbf{1} & \cellcolor{applegreen!30}\textbf{7.84} (10.52) & 3 & 3.95 (1.60) & 4 & 3.90 (2.00) & 2 & 3.97 (2.01) \\
    \textsc{RPCS3} & \cellcolor{applegreen!30}\textbf{1} & \cellcolor{applegreen!30}\textbf{4.44} (8.54) & 2 & 2.39 (7.72) & 3 & 0.77 (5.19) & \textbf{1} & \textbf{3.52} (6.46) \\
    \textsc{Faster‑Whisper} & \cellcolor{applegreen!30}\textbf{1} & \cellcolor{applegreen!30}\textbf{5.64} (3.58) & 4 & 1.58 (5.78) & 3 & 4.08 (2.91) & 2 & 4.14 (1.66) \\
    \textsc{Langflow} & \cellcolor{applegreen!30}\textbf{1} & \cellcolor{applegreen!30}\textbf{9.01} (2.33) & 2 & 2.72 (2.26) & 4 & 1.63 (1.68) & 3 & 2.34 (2.61) \\
    \hline
        Average $r$ & \multicolumn{2}{l}{\textbf{1.00}} & \multicolumn{2}{l}{2.40} & \multicolumn{2}{l}{3.00} & \multicolumn{2}{l}{2.00}
    \\
    \bottomrule
    \end{tabular}
    \end{adjustbox}
    \end{table}

%% file: Tables/RQ2_results.tex
\begin{table}[t!]
    \footnotesize
    \centering
    \caption{The mean and standard deviation of \%PI, denoted as Mean (SD), for \model~and its ablated versions across three LLMs and five projects. For each case, \setlength{\fboxsep}{1.5pt}\colorbox{applegreen!30}{green cells} mean \model~has the best mean \%PI; or \setlength{\fboxsep}{1.5pt}\colorbox{red!20}{red cells} otherwise. The one(s) with the best rank ($r$) from the statistical tests is in bold.}
    \label{tab:rq2_results}
    \setlength{\tabcolsep}{0.6mm}
    \begin{adjustbox}{width=\columnwidth,center}
    \begin{tabular}{lllllllll}
    \toprule
    \multirow{2}{*}{\textbf{System}} & \multicolumn{2}{c}{\textbf{\model}} & \multicolumn{2}{c}{\textbf{\model$_{\texttt{NP}}$}} & \multicolumn{2}{c}{\textbf{\model$_{\texttt{NT}}$}} & \multicolumn{2}{c}{\textbf{\model$_{\texttt{NL}}$}} \\ \cline{2-9}
 & \textbf{$r$} & \textbf{Mean (SD)} & \textbf{$r$} & \textbf{Mean (SD)} & \textbf{$r$} & \textbf{Mean (SD)} & \textbf{$r$} & \textbf{Mean (SD)} \\ \hline
    \textsc{BitmapPlusPlus} & \textbf{1} & \textbf{19.06} (1.48) & \cellcolor{red!20}\textbf{1} & \cellcolor{red!20}\textbf{19.18} (2.22) & \textbf{1} & \textbf{19.14} (2.01) & \textbf{1} & \textbf{19.11} (1.95) \\
    \textsc{Llama.cpp} & \cellcolor{applegreen!30}\textbf{1} & \cellcolor{applegreen!30}\textbf{7.84} (10.52) & 4 & 1.85 (0.48) & 2 & 2.10 (1.59) & 3 & 1.99 (0.56) \\
    \textsc{RPCS3} & \cellcolor{applegreen!30}\textbf{1} & \cellcolor{applegreen!30}\textbf{4.44} (8.54) & \textbf{1} & \textbf{4.30} (9.39) & 2 & 2.45 (7.49) & \textbf{1} & \textbf{3.82} (10.31) \\
    \textsc{Faster‑Whisper} & \textbf{1} & \textbf{5.64} (3.58) & 3 & -0.37 (2.88) & 2 & -0.36 (6.48) & \cellcolor{red!20}\textbf{1} & \cellcolor{red!20}\textbf{5.73} (1.76) \\
    \textsc{Langflow} & \cellcolor{applegreen!30}\textbf{1} & \cellcolor{applegreen!30}\textbf{9.01} (2.33) & 2 & 2.03 (3.21) & 3 & 1.87 (2.82) & 4 & 1.22 (1.02) \\
    \hline
        Average $r$ & \multicolumn{2}{l}{\textbf{1.00}} & \multicolumn{2}{l}{2.20} & \multicolumn{2}{l}{2.00} & \multicolumn{2}{l}{2.00}
    \\
    \bottomrule
    \end{tabular}
    \end{adjustbox}
    \end{table}

%% file: Tables/RQ3_results.tex
\begin{table}[t!]
    \footnotesize
    \centering
    \caption{The mean and standard deviation of \%PI, denoted as Mean (SD), for different meta-prompting LLMs across five projects. For each case, \setlength{\fboxsep}{1.5pt}\colorbox{applegreen!30}{green cells} mean the approach has the best mean \%PI. The one(s) with the best rank ($r$) from the statistical tests is in bold.}
    \label{tab:rq3_results}
    \setlength{\tabcolsep}{0.6mm}
    \begin{adjustbox}{width=\columnwidth,center}
    \begin{tabular}{lllllll}
    \toprule
    \multirow{2}{*}{\textbf{System}} & \multicolumn{2}{c}{\textbf{\model$_{\texttt{4o}}$}} & \multicolumn{2}{c}{\textbf{\model$_{\texttt{37}}$}} & \multicolumn{2}{c}{\textbf{\model$_{\texttt{25}}$}} \\ \cline{2-7}
 & \textbf{$r$} & \textbf{Mean (SD)} & \textbf{$r$} & \textbf{Mean (SD)} & \textbf{$r$} & \textbf{Mean (SD)} \\ \hline
    \textsc{BitmapPlusPlus} & \cellcolor{applegreen!30}\textbf{1} & \cellcolor{applegreen!30}\textbf{19.77} (2.01) & 2 & 19.30 (1.39) & \textbf{1} & \textbf{19.50} (1.85) \\
    \textsc{Llama.cpp} & \textbf{1} & \textbf{5.30} (0.33) & 2 & 4.97 (0.99) & \cellcolor{applegreen!30}\textbf{1} & \cellcolor{applegreen!30}\textbf{5.38} (0.40) \\
    \textsc{RPCS3} & \textbf{1} & \textbf{0.40} (1.11) & 2 & 0.19 (0.26) & \cellcolor{applegreen!30}\textbf{1} & \cellcolor{applegreen!30}\textbf{0.63} (1.25) \\
    \textsc{Faster‑Whisper} & \cellcolor{applegreen!30}\textbf{1} & \cellcolor{applegreen!30}\textbf{3.17} (3.74) & 3 & 0.20 (6.92) & 2 & 2.06 (3.58) \\
    \textsc{Langflow} & \cellcolor{applegreen!30}\textbf{1} & \cellcolor{applegreen!30}\textbf{9.64} (3.32) & 2 & 8.60 (1.67) & \textbf{1} & \textbf{9.31} (1.93) \\
    \hline
        Average $r$ & \multicolumn{2}{l}{\textbf{1.00}} & \multicolumn{2}{l}{2.20} & \multicolumn{2}{l}{1.20}
    \\
    \bottomrule
    \end{tabular}
    \end{adjustbox}
    \end{table}

%% file: Sections/discussion.tex
\section{Discussion}
\label{sec:discussion}

\subsection{Practical Validity of \model~Performance Improvements}

Our evaluation across five real-world systems demonstrates that \model's performance improvements are both significant and robust. However, a critical concern in automated code optimization is whether performance improvements represent genuine algorithmic enhancements rather than just removing the original functions. 

To understand the nature of \model's optimizations, we analyzed the optimizations with the top-10 code optimization for each subject system (50 analyses in total). Specifically, their distribution of optimization types is summarized in Figure~\ref{fig:optimization_types}.

Notably, Loop \& Vectorization (26\%), Algorithmic Improvements (20\%), and Code Quality \& Correctness improvements (20\%) are the most targeted optimizations, whereas the small proportion (4\%) of trivial optimizations, such as reducing the number of iterations and adjusting line breaks, indicates that \model~effectively identifies common performance bottlenecks \cite{DBLP:conf/iclr/ShypulaMZ0GYHNR24, gong2025language} and applies appropriate optimization strategies.

\input{Figures/pie_optimization_types}

\subsection{Illustrative Examples of Generated Meta-Prompts}
\label{subsec:prompt_examples}

To illustrate \model's superiority over baseline methods, we examine concrete examples from \textsc{Llama.cpp}, where \model~achieved 97.5\% higher \%PI over the best baseline (7.87 vs 3.97 in Table~\ref{tab:rq1_results}). 

Particularly, \model~generates prompts with comprehensive optimization strategies, while adapting to each LLM's strengths (Appendix~\ref{sec:appendix}). For example: GPT-4o (Figure~\ref{fig:mpco_gpt4o_prompt}) focuses on ``complex interdependencies and comprehensive optimization''; Claude 3.7 Sonnet (Figure~\ref{fig:mpco_claude_prompt}) promotes ``systematic architectural thinking with maintainability considerations; and Gemini 2.5 Pro (Figure~\ref{fig:mpco_gemini_prompt}) emphasizes ``complex reasoning to verify performance metrics''.

In contrast, baseline methods provide generic guidance: chain-of-thought offers step-by-step frameworks without domain expertise (Figure~\ref{fig:cot_prompt}); few-shot learning provides simple examples insufficient for C/C++ optimization (Figure~\ref{fig:fewshot_prompt}); and contextual prompting uses all available contexts (project, task, LLM) uniformly across all cases (Figure~\ref{fig:contextual_prompt}).

Therefore, the key advantage of \model~lies in its dual adaptation strategy: (1) \textbf{task-specific guidance} that addresses the unique characteristics of \textsc{Llama.cpp} (C/C++ codebase, performance-critical AI inference, memory-intensive operations), and (2) \textbf{LLM-specific instructions} tailored to each model's cognitive strengths and architectural capabilities. This explains why \model~achieved top-ranked performance across all systems in \textbf{RQ1}, while baseline methods failed to consistently leverage each LLM's optimization capabilities or address task-specific performance bottlenecks.

\subsection{Meta-Prompting Overhead Analysis}
\label{subsec:overhead_analysis}

\begin{figure}[t!]
    \centering
    \includegraphics[width=\columnwidth]{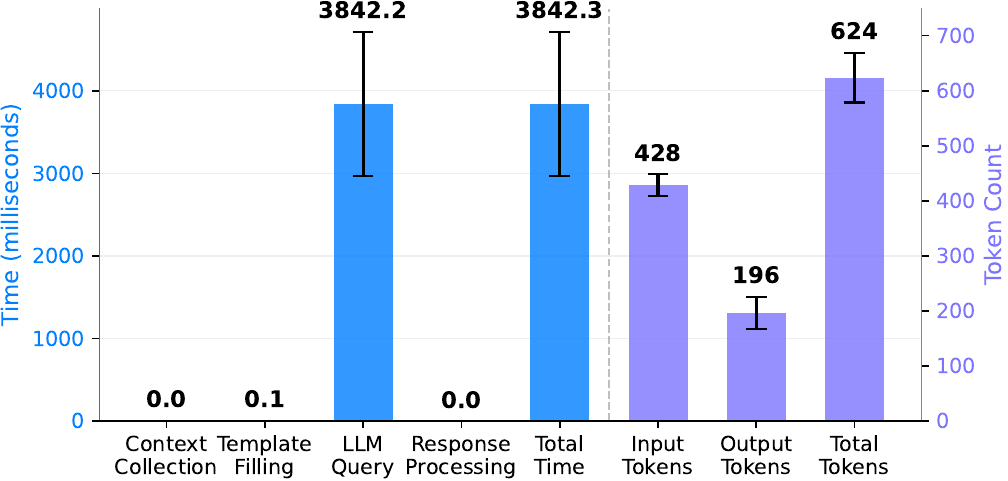} 
    \caption{The time and token overhead of \model.}
    \label{fig:overhead}
\end{figure}

To validate \model's industrial efficiency claims, we conducted a comprehensive overhead analysis by measuring meta-prompting overhead across 50 constructs from all five subject systems, using GPT-4o as the meta-prompter. Figure~\ref{fig:overhead} presents the detailed timing breakdown and token usage analysis.

The results demonstrate that \model~introduces minimal computational overhead: averaging 3.8 seconds per meta-prompt generation with 624 tokens (428 input, 196 output). In particular, the LLM query dominates the timing, while context collection, template filling, and response processing contribute negligible overhead ($<$0.1ms each). At GPT-4o pricing (\$2.5/1M input tokens, \$10/1M output tokens), each meta-prompt generation costs approximately \$0.003, making the approach highly practical for industrial deployment.

Importantly, \model~provides substantial cost advantages over traditional prompt engineering approaches. Manual prompt crafting requires expert engineers to spend minutes to hours developing and tuning LLM-specific prompts for each optimization task, followed by ongoing maintenance as LLMs evolve. In contrast, \model~automates this entire process at \$0.003 per LLM-task pair, orders of magnitude lower cost than evolutionary prompt optimization or iterative agent-based methods that require multiple LLM calls. The single-call design ensures that once generated, meta-prompts can be reused indefinitely without additional overhead, reducing both the expertise requirement and maintenance burden while delivering superior optimization effectiveness across diverse tasks and LLMs.

\subsection{Actionable Guidelines for Context Provision}

Our ablation study in Section~\ref{subsec:rq2} offers clear insights into how context should be structured for effective meta-prompting in industrial environments. Particularly, the full \model~template achieved the best average rank across all systems, while removing any contextual component noticeably reduces effectiveness. This demonstrates that \model~performs best when supplied with comprehensive information, including target LLM capabilities, optimization objectives and considerations, and project metadata. Therefore, we have:

\keybox{
   \textit{\textbf{Suggestion 1:} Maintain comprehensive context integration (project, task, and LLM context), as removing any contextual component might significantly degrade performance, with project context removal causing the highest degradation.}
}

In our implementation of \model, we developed a modular context-integration pipeline that queries the platform’s existing APIs to retrieve contextual data from the database. By leveraging native services, we avoided duplicating functionality and maintained low-overhead data access. Furthermore, the contexts were stored using a lightweight, structured JSON schema that can be incrementally extended and updated as projects evolve, without requiring extensive re-engineering of the prompting infrastructure. As such:

\keybox{
   \textit{\textbf{Suggestion 2:} Use a modular, API-driven pipeline that integrates with existing platform services to enable rapid context-collection, and represent context in a structured schema to support flexible, incremental extension as the project evolves.}
}

\subsection{Actionable Guidelines for Meta-Prompting LLM Selection}
The comprehensive evaluation across multiple LLM configurations in Section~\ref{subsec:rq3} provides practical recommendations for organizations deploying \model~in industrial environments. They show that all three major LLMs can serve effectively as meta-prompters, with GPT-4o achieving the best overall performance (average rank of 1.00) and Gemini 2.5 Pro also highly competitive (average rank of 1.20). Nevertheless, since the performance differences are relatively small, platforms can confidently base their meta-prompting LLM choice on pragmatic factors such as cost constraints, model availability, and integration requirements without compromising \model's effectiveness. In summary:

\keybox{
   \textit{\textbf{Suggestion 3:} Prioritize deployment considerations (cost, availability, integration) over minor performance differences when selecting meta-prompting LLMs, as all three major providers offer comparable effectiveness.}
}

Beyond the current one-step meta-prompting, platforms could also explore advanced meta-prompting techniques to enhance optimization effectiveness. For example, agent-based meta-prompting, where multiple specialized agents collaborate to dynamically collect contexts based on the task \cite{wang2023promptagent}, and iterative meta-prompting, where prompts are refined through multiple rounds of prompt engineering \cite{zhang2023meta}. While incurring additional computation and coordination overhead, these techniques can handle more complicated optimization scenarios that involve multiple interdependent code blocks or require dynamic and domain-specific contexts. Based on the above, we have:

\keybox{
   \textit{\textbf{Suggestion 4:} Consider advanced meta-prompting techniques (agent-based, iterative) as alternatives to one-step meta-prompting when platforms want to trade off computational resources for enhanced optimization performance.}
}

\subsection{Limitations and Future Work}

While \model~demonstrates significant effectiveness in addressing the cross-model prompt engineering bottleneck, several limitations warrant discussion. 

\textbf{Language and domain:} Our evaluation focuses on C++ and Python in specific domains (HPC, quantitative finance, data processing), and generalization to other languages and domains requires further validation. {Future work could evaluate the effectiveness of meta-prompting across diverse programming languages and application domains to establish broader generalizability.}

\textbf{Optimization granularity:} Currently, we only address function-level performance bottlenecks rather than file-level or system-level optimizations that require coordinated changes across multiple functions or entire systems. {Future work could extend \model~to handle multi-level optimization scenarios, developing context collection strategies that capture inter-function dependencies and system-wide performance characteristics.}

\textbf{Optimization target:} This study targets runtime optimization and validation, without exploring other objectives like CPU usage, memory usage, and energy efficiency. {Future work could easily expand \model's capabilities to handle different metric optimization scenarios, or even develop multi-objective meta-prompting strategies that can balance competing optimization goals.}

\textbf{Verification reliability:} Our validation of functional correctness relies on existing unit test suites of target projects---if original test coverage is low, LLM-generated optimized code might introduce subtle bugs or edge case failures that go undetected. Future work should investigate integrating automatic test case generation, property-based testing, or symbolic checks to create more robust verification pipelines that are less dependent on existing test coverage.


%% file: Figures/pie_optimization_types.tex
\begin{figure}[t]
\centering
\begin{tikzpicture}
  \def\center{0,0}
  \def\radius{1.8cm}

  \def\linelength{0.4cm}
  \def\textdistance{2.1cm}

  \def\colorloopvec{beaublue}
  \def\coloralgorithmic{cerublue!80}
  \def\colordatamemory{applegreen}
  \def\colorcache{frenchblue!80}
  \def\colorparallel{ao!60}
  \def\colorquality{iceberg}
  \def\colorinvalid{amaranth}

  \def\loopvec{26}
  \def\algorithmic{20}
  \def\datamemory{16}
  \def\cache{8}
  \def\parallel{6}
  \def\quality{20}
  \def\invalid{4}

  \def\total{100}
  \def\loopvecangle{360*\loopvec/\total}
  \def\algorithmicangle{360*\algorithmic/\total}
  \def\datamemoryangle{360*\datamemory/\total}
  \def\cacheangle{360*\cache/\total}
  \def\parallelangle{360*\parallel/\total}
  \def\qualityangle{360*\quality/\total}
  \def\invalidangle{360*\invalid/\total}

  \fill[\colorloopvec]  (0,0) -- (\loopvecangle:\radius)                                                                                                          arc (\loopvecangle:0:\radius) -- cycle;
  \fill[\coloralgorithmic](0,0) -- (\loopvecangle+\algorithmicangle:\radius)                                                                                   arc (\loopvecangle+\algorithmicangle:\loopvecangle:\radius) -- cycle;
  \fill[\colordatamemory](0,0) -- (\loopvecangle+\algorithmicangle+\datamemoryangle:\radius)                                                                    arc (\loopvecangle+\algorithmicangle+\datamemoryangle:\loopvecangle+\algorithmicangle:\radius) -- cycle;
  \fill[\colorcache]     (0,0) -- (\loopvecangle+\algorithmicangle+\datamemoryangle+\cacheangle:\radius)                                                        arc (\loopvecangle+\algorithmicangle+\datamemoryangle+\cacheangle:\loopvecangle+\algorithmicangle+\datamemoryangle:\radius) -- cycle;
  \fill[\colorparallel]  (0,0) -- (\loopvecangle+\algorithmicangle+\datamemoryangle+\cacheangle+\parallelangle:\radius)                                          arc (\loopvecangle+\algorithmicangle+\datamemoryangle+\cacheangle+\parallelangle:\loopvecangle+\algorithmicangle+\datamemoryangle+\cacheangle:\radius) -- cycle;
  \fill[\colorquality]   (0,0) -- (\loopvecangle+\algorithmicangle+\datamemoryangle+\cacheangle+\parallelangle+\qualityangle:\radius)                           arc (\loopvecangle+\algorithmicangle+\datamemoryangle+\cacheangle+\parallelangle+\qualityangle:\loopvecangle+\algorithmicangle+\datamemoryangle+\cacheangle+\parallelangle:\radius) -- cycle;
  \fill[\colorinvalid]   (0,0) -- (\loopvecangle+\algorithmicangle+\datamemoryangle+\cacheangle+\parallelangle+\qualityangle+\invalidangle:\radius)             arc (\loopvecangle+\algorithmicangle+\datamemoryangle+\cacheangle+\parallelangle+\qualityangle+\invalidangle:\loopvecangle+\algorithmicangle+\datamemoryangle+\cacheangle+\parallelangle+\qualityangle:\radius) -- cycle;

  \node at (\loopvecangle/2:0.7*\radius)                                {\small 26\%};
  \node at (\loopvecangle+\algorithmicangle/2:0.7*\radius)              {\small 20\%};
  \node at (\loopvecangle+\algorithmicangle+\datamemoryangle/2:0.7*\radius) {\small 16\%};
  \node at (\loopvecangle+\algorithmicangle+\datamemoryangle+\cacheangle/2:0.7*\radius)  {\small 8\%};
  \node at (\loopvecangle+\algorithmicangle+\datamemoryangle+\cacheangle+\parallelangle/2:0.7*\radius) {\small 6\%};
  \node at (\loopvecangle+\algorithmicangle+\datamemoryangle+\cacheangle+\parallelangle+\qualityangle/2:0.7*\radius) {\small 20\%};
  \node at (\loopvecangle+\algorithmicangle+\datamemoryangle+\cacheangle+\parallelangle+\qualityangle+\invalidangle/2:0.7*\radius) {\small 4\%};

  \draw[\colorloopvec, thick] (\loopvecangle/2:\radius) -- (\loopvecangle/2:\radius+\linelength);
  \node[anchor=west, align=left] at (\loopvecangle/2:\textdistance)
    {\small Loop \\ \& Vectorization};

  \draw[\coloralgorithmic, thick] (\loopvecangle+\algorithmicangle/2:\radius) -- (\loopvecangle+\algorithmicangle/2:\radius+\linelength);
  \node[anchor=south, align=left] at (\loopvecangle+\algorithmicangle/2:\textdistance)
    {\small Algorithmic \\ Improvements};

  \draw[\colordatamemory, thick] (\loopvecangle+\algorithmicangle+\datamemoryangle/2:\radius) -- (\loopvecangle+\algorithmicangle+\datamemoryangle/2:\radius+\linelength);
  \node[anchor=east, align=left] at (\loopvecangle+\algorithmicangle+\datamemoryangle/2:\textdistance)
    {\small Data Structure \\ \& Memory};

  \draw[\colorcache, thick] (\loopvecangle+\algorithmicangle+\datamemoryangle+\cacheangle/2:\radius) -- (\loopvecangle+\algorithmicangle+\datamemoryangle+\cacheangle/2:\radius+0.6cm);
  \node[anchor=north east, align=left] at (\loopvecangle+\algorithmicangle+\datamemoryangle+\cacheangle/2:\textdistance)
    {\small Caching \& \\ Memoization};

  \draw[\colorparallel, thick] (\loopvecangle+\algorithmicangle+\datamemoryangle+\cacheangle+\parallelangle/2:\radius) -- (\loopvecangle+\algorithmicangle+\datamemoryangle+\cacheangle+\parallelangle/2:\radius+\linelength);
  \node[anchor=north west, align=left] at (\loopvecangle+\algorithmicangle+\datamemoryangle+\cacheangle+\parallelangle/2:\textdistance)
    {\small Parallelization \\ \& Concurrency};

  \draw[\colorquality, thick] (\loopvecangle+\algorithmicangle+\datamemoryangle+\cacheangle+\parallelangle+\qualityangle/2:\radius) -- (\loopvecangle+\algorithmicangle+\datamemoryangle+\cacheangle+\parallelangle+\qualityangle/2:\radius+\linelength);
  \node[anchor=west, align=left] at (\loopvecangle+\algorithmicangle+\datamemoryangle+\cacheangle+\parallelangle+\qualityangle/2:\textdistance)
    {\small Code Quality \& \\ Correctness};

  \draw[\colorinvalid, thick] (\loopvecangle+\algorithmicangle+\datamemoryangle+\cacheangle+\parallelangle+\qualityangle+\invalidangle/2:\radius) -- (\loopvecangle+\algorithmicangle+\datamemoryangle+\cacheangle+\parallelangle+\qualityangle+\invalidangle/2:\radius+\linelength);
  \node[anchor=west, align=left] at (\loopvecangle+\algorithmicangle+\datamemoryangle+\cacheangle+\parallelangle+\qualityangle+\invalidangle/2:\textdistance)
    {\small Trivial};

\end{tikzpicture}
\caption{Distribution of optimization types among the top 10\% performing optimizations generated by \model.}
\label{fig:optimization_types}
\end{figure}

%% file: Sections/conclusion.tex
\section{Conclusion}
\label{sec:conclusion}
We have presented \model, a meta-prompting approach that automatically optimizes and selects effective prompts for code optimization across multiple LLMs. Over 366 hours of benchmarking, \model~demonstrated significant performance gains on five real-world codebases while meeting industrial efficiency requirements. This makes it practical for automated code optimization platforms such as \textsc{Artemis} to adopt multi-LLM optimization strategies without the burden of manual prompt engineering.  

Based on our analysis, we recommend maintaining comprehensive context integration, automating context collection using existing infrastructure, selecting meta-prompting LLMs according to deployment constraints rather than marginal model differences, and applying advanced techniques only when higher performance justifies the additional computational cost.  Our future work will extend \model~to support multi-level and multi-objective optimization scenarios, as well as integrating automatic test generation for more robust verification pipelines.

\section*{Acknowledgment}
This work was supported in part by an Innovative UK Knowledge Transfer Partnership (KTP) project between the University of Leeds and TurinTech AI, and the UK Engineering
and Physical Sciences Research Council (EPSRC) under
grant agreement EP/X037304/1.





%% file: Sections/appendix.tex
\section{Appendix: Prompt Examples}
\label{sec:appendix}



\begin{figure}[h!]
    \centering
    \begin{lstlisting}[basicstyle=\footnotesize\ttfamily, frame=single, breaklines=true, xleftmargin=0.2cm, xrightmargin=0.2cm, moredelim={[is][\color{blue}]{@}{@}}, moredelim={[is][\color{applegreen}]{§}{§}}, moredelim={[is][\color{frenchblue}]{¤}{¤}}, moredelim={[is][\color{amaranth}]{€}{€}}]
Optimize the C code in the llama.cpp project to enhance runtime performance by focusing on the following key areas:

@1. **Algorithmic Complexity**: Analyze the code to identify and reduce the Big O complexity where possible. Consider alternative algorithms that offer better performance.
2. **Data Structure Efficiency**: Evaluate the current data structures for efficiency in terms of access patterns and memory usage. Replace or modify them to improve performance.
3. **Loop Optimizations**: Inspect loops for unnecessary iterations and optimize them. Consider loop unrolling and other techniques to reduce overhead.
4. **Memory Access Patterns**: Improve memory access patterns to enhance cache utilization. Minimize cache misses by optimizing data locality.
5. **I/O Operations**: Reduce the overhead of I/O operations and system calls. Batch operations where possible to minimize latency.
6. **Parallel Processing**: Identify opportunities for parallel processing. Use multithreading or SIMD instructions to leverage multiple cores and vectorization.
7. **Redundant Computations**: Eliminate redundant computations by caching results or reusing previous calculations where applicable.@

§For larger models like gpt-4-o, ensure to internally verify assumptions and metrics based on the task description, and consider complex interdependencies within the code for comprehensive optimization.§
    \end{lstlisting}
    \caption{MPCO generated prompt for runtime performance optimization in Llama.cpp targeting GPT-4o.}
    \label{fig:mpco_gpt4o_prompt}
\end{figure}

\begin{figure}[h!]
    \centering
    \begin{lstlisting}[basicstyle=\footnotesize\ttfamily, frame=single, breaklines=true, xleftmargin=0.2cm, xrightmargin=0.2cm, moredelim={[is][\color{blue}]{@}{@}}, moredelim={[is][\color{applegreen}]{§}{§}}, moredelim={[is][\color{frenchblue}]{¤}{¤}}, moredelim={[is][\color{amaranth}]{€}{€}}]
Optimize the C code in the llama.cpp project to enhance runtime performance through systematic analysis and targeted improvements:

@1. **Algorithmic Complexity**: Examine the fundamental algorithms and identify opportunities to reduce computational complexity. Prioritize algorithmic changes that offer the greatest performance gains.
2. **Data Structure Optimization**: Assess current data structures for access efficiency and memory footprint. Consider restructuring data layouts to improve cache performance and reduce memory overhead.
3. **Loop and Control Flow**: Analyze loop structures for optimization opportunities including unrolling, vectorization, and elimination of unnecessary branching. Focus on hot paths and frequently executed code sections.
4. **Memory Management**: Optimize memory access patterns to maximize cache efficiency. Consider memory prefetching, data alignment, and reducing memory fragmentation.
5. **System-Level Optimizations**: Evaluate I/O operations, system calls, and resource utilization. Implement batching and buffering strategies to minimize system overhead.
6. **Concurrency and Parallelization**: Identify parallelizable sections and implement appropriate threading or SIMD optimizations. Balance parallelization overhead with performance gains.
7. **Code Efficiency**: Eliminate redundant computations, optimize critical path execution, and implement result caching where beneficial.@

§For claude-3-7-sonnet, approach optimization with systematic architectural thinking, considering both micro-optimizations and broader structural improvements. Provide clear rationale for each optimization decision and consider the maintainability implications of proposed changes.§
    \end{lstlisting}
    \caption{MPCO generated prompt for runtime performance optimization in Llama.cpp targeting Claude 3.7 Sonnet.}
    \label{fig:mpco_claude_prompt}
\end{figure}


\begin{figure}[h!]
    \centering
    \begin{lstlisting}[basicstyle=\footnotesize\ttfamily, frame=single, breaklines=true, xleftmargin=0.2cm, xrightmargin=0.2cm, moredelim={[is][\color{blue}]{@}{@}}, moredelim={[is][\color{applegreen}]{§}{§}}, moredelim={[is][\color{frenchblue}]{¤}{¤}}, moredelim={[is][\color{amaranth}]{€}{€}}]
Optimize the C code in the llama.cpp project to enhance runtime performance by addressing the following aspects:

@1. **Algorithmic Complexity**: Analyze and reduce the Big O complexity where possible. Identify and refactor inefficient algorithms.
2. **Data Structures**: Evaluate the efficiency of current data structures. Optimize access patterns and consider alternatives that offer better performance.
3. **Loop Optimizations**: Minimize unnecessary iterations. Unroll loops where beneficial and eliminate redundant calculations within loops.
4. **Memory Access Patterns**: Improve cache utilization by optimizing memory access patterns. Ensure data locality and minimize cache misses.
5. **I/O Operations**: Streamline I/O operations and reduce the frequency of system calls. Buffer data effectively to minimize latency.
6. **Parallel Processing**: Identify sections of the code that can be parallelized. Utilize multi-threading or SIMD instructions to leverage multi-core processors.
7. **Redundant Computations**: Detect and eliminate redundant computations. Cache results of expensive operations when possible to avoid repeated calculations.@

§For the gemini-v25-pro model, apply complex reasoning to verify assumptions about performance metrics and ensure optimizations align with the project's goals.§
    \end{lstlisting}
    \caption{MPCO generated prompt for runtime performance optimization in Llama.cpp targeting Gemini 2.5 Pro.}
    \label{fig:mpco_gemini_prompt}
\end{figure}



\begin{figure}[h!]
    \centering
    \begin{lstlisting}[basicstyle=\footnotesize\ttfamily, frame=single, breaklines=true, xleftmargin=0.2cm, xrightmargin=0.2cm, moredelim={[is][\color{blue}]{@}{@}}, moredelim={[is][\color{applegreen}]{§}{§}}]
Let's optimize the following code step by step:

Please follow these reasoning steps:
@1. First, analyze the current code to identify performance bottlenecks
2. Consider different optimization strategies (algorithmic, data structure, loop optimization, etc.)
3. Evaluate the trade-offs of each approach
4. Select the best optimization strategy
5. Implement the optimized version@

§Think through each step, then provide only the final optimized code.§
    \end{lstlisting}
    \caption{Chain-of-Thought (CoT) prompting method used as baseline.}
    \label{fig:cot_prompt}
\end{figure}


\begin{figure}[h!]
    \centering
    \begin{lstlisting}[basicstyle=\footnotesize\ttfamily, frame=single, breaklines=true, xleftmargin=0.2cm, xrightmargin=0.2cm, moredelim={[is][\color{blue}]{@}{@}}, moredelim={[is][\color{applegreen}]{§}{§}}]
Here are examples of code optimization:

@Example 1 - Loop optimization:
Original: for i in range(len(arr)): if arr[i] > threshold: result.append(arr[i])
Optimized: result = [x for x in arr if x > threshold]

Example 2 - Algorithm optimization:
Original: for i in range(n): for j in range(n): if matrix[i][j] > 0: count += 1
Optimized: count = np.sum(matrix > 0)

Example 3 - Data structure optimization:
Original: items = []; for x in data: items.append(x); return sorted(items)
Optimized: return sorted(data)@

§Now optimize the code for better runtime performance, then provide only the final optimized code.§
    \end{lstlisting}
    \caption{Few-shot learning prompting method used as baseline.}
    \label{fig:fewshot_prompt}
\end{figure}


\begin{figure}[h!]
    \centering
    \begin{lstlisting}[basicstyle=\footnotesize\ttfamily, frame=single, breaklines=true, xleftmargin=0.2cm, xrightmargin=0.2cm, moredelim={[is][\color{blue}]{@}{@}}, moredelim={[is][\color{applegreen}]{§}{§}}, moredelim={[is][\color{frenchblue}]{¤}{¤}}, moredelim={[is][\color{amaranth}]{€}{€}}]
You are an expert in code optimization. Please optimize the provided code for @{objective}@. Consider the project context, task context, and adapt your optimization approach accordingly.
§## Project Context
Project Name: {project_name}
Project Description: {project_description}
Primary Languages: {project_languages}§

¤## Task Context
- Description: {task_description}
- Considerations: {task_considerations}¤

€## Target LLM Context
- Target Model: {target_llm}
- Considerations: {llm_considerations}€
    \end{lstlisting}
    \caption{Contextual prompting template used as baseline method, which uses all available contexts but without meta-prompting adaptation (the context placeholders will be filled in real prompts).}
    \label{fig:contextual_prompt}
\end{figure}